\documentclass[10pt]{article}
\usepackage{amsfonts,amsmath,amssymb,dsfont}
\usepackage[left=2cm,right=2cm,top=2cm,bottom=2cm,bindingoffset=0cm]{geometry}
\usepackage[normalem]{ulem}
\usepackage[natbib=true, backend = bibtex, style=numeric-comp, sorting=none,maxbibnames=99]{biblatex}
\usepackage{graphicx}
\def\be{\begin{eqnarray}}
\def\ee{\end{eqnarray}}
\def\nn{\nonumber}

\def\Rm{\mathcal{R}}

\author{A.Anokhina\footnote{Kurchatov Institute, Moscow; Kharkevich Institute for Information Transmission Problems, Moscow; anokhina@itep.ru}}
\title{Towards Catastrophe theory for Khovanov--Rozansky homology}
\date{}

\begin{document}

\maketitle

\vspace{-5cm}

\begin{flushright}
ITEP-TH-7/24\\
IITP-TH-6/24
\end{flushright}

\vspace{3cm}

\begin{abstract}
We briefly summarise our results on jumps in the analytic formulas for the Khovanov(--Rozansky) polynomials. We conclude from the empiric data that there are ``regular'' and ``weird'' catastrophes, which drastically differ by form of the associated jumps in the Khovanov(--Rozansky) polynomials. This is the first step towards the catastrophe theory for the cohomological knot invariants. In particular, it can be another way to see these quantities as observables in cohomological quantum field theory. 
\end{abstract}
\section{Introduction}
Catastrophe theory \cite{Arn} is a powerful tool of mathematical physics to study non-linear dynamical systems \cite{MandMD}. It commonly operates with systems of differential equations, but is close in spirit to homological calculus in topology \cite{GelMan}. A kind of hybrid of the two fields is known as cohomological quantum field theory (CQFT) \cite{GukSt,GSchV,Gal,AnKh}. Such models seem to be very interesting and profound, and they might be useful in various applications as new tools of the Catastrophe theory.  
Our goal is to use the knot homology calculus to develop a ``Cohomological Catastrophe theory''. By this we mean a family of constructively defined CQFT models related by a kind of evolution ``regular'' on the moduli space except for special (``catastrophe'') points.
In this letter we summarise our current knowledge on the Catastrophes for explicitly studied CQFT models associated with the Khovanov--Rozansky homology \cite{Khov,BarNat,KhR,CarMuf} of several knot families \cite{AMev,APev,ALMSat}. 

\section{Two basic kinds of $KhR_N$ catastrophes}
The  $SU_N$ Khovanov--Rozhansky polynomial ($KhR_N$) is labelled by the gauge group rank $N$, depends on the formal ``quantum'' variable $q$ and on the formal ``homological'' variable $t$, and has the HOMFLY polynomial as a boundary condition at $t=-1$,
$
KhR_N\big(t=-1,q\big)=H_N(q).
$
The coefficients of the $q$ and $t$ powers in the $KhR$ enumerate the dimensions of homologies and hence are positive integers. Below we call a polynomial with all positive (negative) coefficients a \textit{positive (negative) polynomial}, and we call it a \textit{sign indefinite polynomial} when the signs vary.
 
The $\Rm$-matrix formalism for HOMFLY polynomials \cite{MorSm} implies that HOMFLY polynomials for a knot family generated by \textit{evolution} of the knot diagram, i.e., by inserting somewhere a repeated fragment (e.g., a two strand braid with $2n+1$ crossings) depends analytically on the family parameters $n_i$ through the exponentials $\lambda_i^{n_i}$, where the $\lambda_i$ (which we call \textit{eigenvalues}) are common for the knot family and, moreover, for many knot families \cite{EvoDiff}. Surprisingly, a very similar exponential behaviour along the same knot families takes place for the $KhR_N$ \cite{APev,AMev,ALMSat}, up to separate jumps of analytic expression that due to the boundary condition take the form 
$
KhR_N \to KhR_N+(1+t)KhR_N^{\prime\prime}.
$
Equivalently, one can write
\be
KhR_N=KhR_N^{\prime}+(-t)^{\Theta_s}KhR_N^{\prime\prime},\qquad \Theta_s=\left\{\begin{array}{l}1,\ s\ge0\\
0,\ s<0\end{array}\right.,\label{gencat}
\ee
where $s$ depends on the family parameters $\{n_i\}$ as well as on the original knot  (with $n_i=0$), and both the $KhR_N^{\prime}$ and $KhR_N^{\prime\prime}$ have no jumps at the point where $s=0$. We call such violations of analytic behaviour a \textbf{catastrophe} of the $KhR_N$ (in analogy with a catastrophe of an analytic solution of a non-linear ODE).

\section{Regular catastrophes}
In this section, we consider the cases, where $KhR_N^{\prime\prime}$ in (\ref{gencat}) is proportional to or looks like the $KhR_N$ polynomial for a two-strand knot. 
We call such catastrophes \textit{regular}.
The $KhR_N$ complex for a two-strand braid has almost the same form for all $N$ \cite{CarMuf}. Hence,
we can focus here on the Khovanov polynomial $Kh\equiv KhR_2$, for which most of our explicit formulas are relevant.

\begin{figure}
$$
\begin{array}{cc}
\begin{array}{c}
\includegraphics[width=4cm]{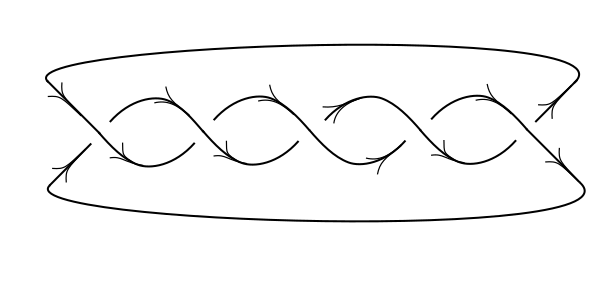}\\[-5mm]
a.\ Tor_{5}\\
\includegraphics[width=4cm]{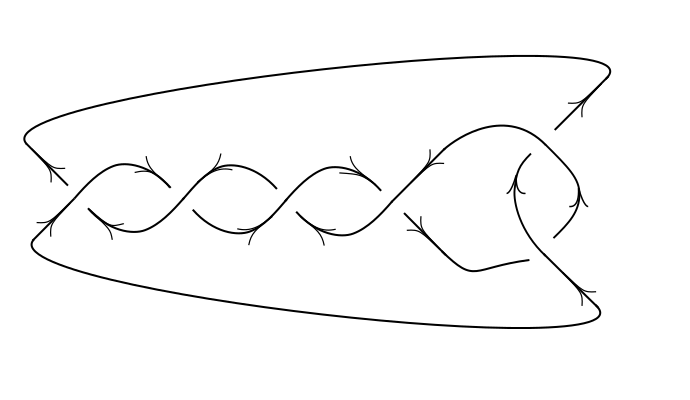}\\[-5mm]
b.\ Tw_{4}
\end{array}
&
\begin{array}{c}
\includegraphics[width=6cm]{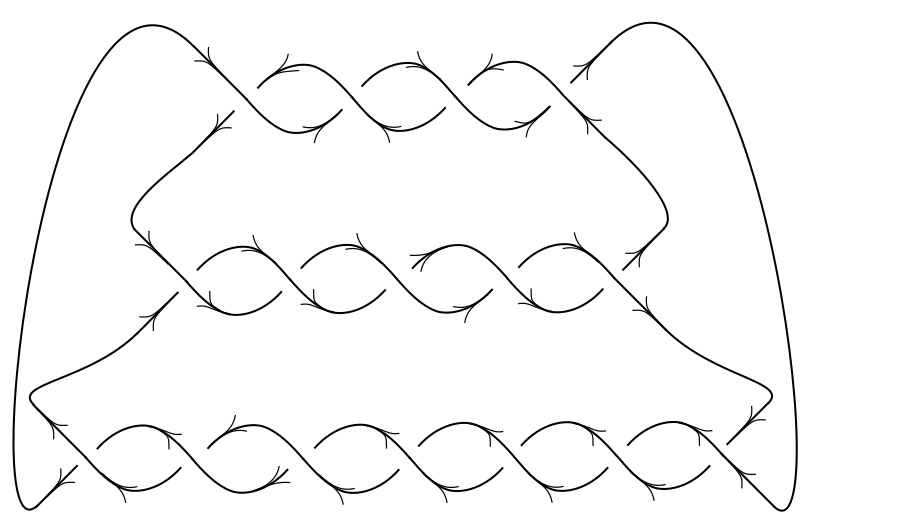}\\
c.\ P^{7,5,-4}
\end{array}
\end{array}
$$
\caption{a,b. Two-strand torus and twist knots. c. Pretzel knot. \label{fig:pretz}}
\end{figure}

\paragraph{Two-strand torus and twist knots.}
The basic ingredient in the examples below is the $Kh$ of the two-strand torus knot \cite{BarNat}, which is the standard closure of the two-strand parallel braid Fig.\ref{fig:pretz}.a,
\be
\begin{array}{c}
Kh^{\mathrm{Tor}_{2,2n+1}}=(-t)^{-\Theta_{-n}}q^{2n}F_{2n+1}(q^2t),\qquad \Theta_n=\left\{\begin{array}{l}1,\ n\ge0\\
0,\ n<0\end{array}\right.,\\\\
F_{2n+1}(\lambda)= 1+\lambda^2\frac{1-\lambda^{2n}}{1-\lambda}=
\left\{\begin{array}{c}
1+\lambda^2+\lambda^3+\ldots+\lambda^{2n+1},\ \ n\ge 0\\
-\lambda-\lambda^{-1}-\lambda^{-2}-\ldots-\lambda^{2-2n}\ \ n<0
\end{array}
\right..\end{array}
\label{KhTor}\ee
For $n>0$, $F_{2n+1}(\lambda)$ is a positive polynomial in $\lambda$, whose coefficients are the dimensions of the $Kh$ homologies. For $n<0$, $F_{2n+1}(\lambda)$ is a negative polynomial, whose coefficients cannot be dimensions of homologies. Yet the factor of  $(-t)^{-\Theta_{-n}}$ changes its value from $1$ for $n>0$ to $-t^{-1}$ for $n<0$ and compensates the signs\footnote{As a price, all the $t$ powers are reduced by $1$, which corresponds to shifting the sequence of homology dimensions by one w.r.t. the sequence of the spaces in the complex.}. The boundary condition for the $Kh$ is preserved since $(-t)^{\Theta_{-n}}\equiv1$ for $t=-1$. As a straightforward check shows, (\ref{KhTor}) respects the mirror symmetry of the Khovanov polynomial \cite{Khov}, namely, 
$Kh_2^{\mathrm{Tor}_{2,2n+1}}(q,t)=Kh_2^{\mathrm{Tor}_{2,-2n-1}}\big(\frac{1}{q},\frac{1}{t}\big)$.

A twist knot, which is the lock-element closure of the two-strand antiparallel braid (Fig.\ref{fig:pretz}.b), has the $Kh$ polynomial
\be
Kh^{\mathrm{Tw}_{2n}}=(-t)^{-\Theta_n}(q^2t)^{-2n}G_{2n}(q^2t),\qquad G_{2n}(\lambda)= 1+\lambda\frac{\left(1+\lambda^2\right)\left(1-\lambda^{2n}\right)}{1-\lambda},\label{KhTwist}
\ee
similar to that of a two-strand torus knot\footnote{Note that the mirror image of a twist knot has also the mirror lock element (Fig.\ref{fig:pretz}.b), and hence $Tw_{-2n}$ is \textit{not} the mirror image of $Tw_{2n}$ \cite{AMKhov}. In \cite{katlas}, knot $(2n)_1$ is the mirror image of $Tw_{-2n}$, and knot $(2n-1)_2$ is topologically equivalent to $Tw_{2n}$.}. Moreover, one can consider a two-parametric hybrid of two-strand torus and twist knots a similar way \cite{DBPev}.

\paragraph{Multiple two-strand braids in ``thin'' pretzel knots.}
The next case is a genus $g$ pretzel knot $P_{n_0,\ldots,n_g}$ (Fig.\ref{fig:pretz}.c).
As demonstrated in \cite{APev}, there is an analytic formula for the $Kh$ polynomial of $P_{n_0,\ldots,n_g}$, when all $n_i>0$. 
Then there is a set of regions in the $\{n_i\}$ parameter space called regular with some $n_i<0$ and $|n_i|$ is enough large, where the Khovanov polynomials satisfy 
\be
Kh^{P_{n_0,\ldots,n_g}}=(-t)^aKh^{P_{n_0>0,\ldots,n_g>0}}
\ee
for some integer $a$ that depends on the region. 
In particular, in the most studied case of genus two pretzel knot, the $Kh$ polynomial in the regular regions is given by
\be
Kh^{P_{n_0,n_1,n_2}}=(-t)^aq^3\Phi_{n_0,n_1,n_2}(q^2t),\qquad 
\Phi_{n_0,n_1,n_2}(\lambda)=\lambda^{n_0+n_1}
\left\{F_{n_0}\big(\textstyle{\frac{1}{\lambda}}\big)F_{n_1}\big(\textstyle{\frac{1}{\lambda}}\big)+
F''_{n_0,n_1}\big(\textstyle{\frac{1}{\lambda}}\big)F'_{n_2}(\lambda)\right\},
\label{KhPretz}\ee
where $F'$ and $F''$ are slightly modified variants of the $F$ from (\ref{KhTor}),
\be
\begin{array}{c}F_{n}(\lambda)= 1+\lambda^2\frac{1-\lambda^{n-1}}{1-\lambda},\qquad
F'_n(\lambda)=\lambda^{-1}F_{n-1}+\lambda^n,\qquad
F''_{n_0,n_1}(\lambda)= 1+\lambda+\lambda^2\frac{1-\lambda^{n_0-1}}{1-\lambda}+\lambda^2\frac{1-\lambda^{n_1-1}}{1-\lambda}.\end{array}
\label{KhPretzPos}\ee
Up to cyclic permutations of the pretzel handles and rotation of the projection plane by $\pi$ (in Fig.\ref{fig:pretz}.c), we can set $n_0\ge n_1\ge n_2$. Then (\ref{KhPretz}) is valid for $n_0\ge n_1\ge n_2>0$ (with $a=0$), $n_0\ge n_1> 0> -n_1> n_2$ (with $a=-1$), $n_0>-n_1>0> n_1\ge n_2$ (with $a=-1$), $0>n_0\ge n_1\ge n_2$ (with $a=-2$)\footnote{The codimension 1 regions with $n_i=0$ or $n_i=\pm1$ must be considered separately.}.

Polynomial (\ref{KhPretz}) is explicitly positive for $n_0>n_1>n_2>1$. It is actually positive or negative in all union of the regular regions, what can be seen from equivalent forms of (\ref{KhPretzPos}). The factor $(-t)^a$ in (\ref{KhPretz}) then makes the resulting $Kh$ polynomial positive.

All pretzel knots in these regions are ``homologically thin'', i.e., their $KhR_N$ polynomials are obtained from their HOMFLY polynomials by substitution $q^2 \to -tq^2$, $q^{2N}\rightarrow -tq^{2N}$ \cite{APev}. In particular, this implies an analytic dependence of the $KhR_N$ on $N$.

\begin{figure}
	$$
	\arraycolsep=1cm
	\begin{array}{cc}
	\includegraphics[width=4cm]{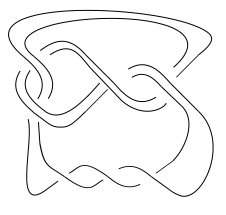}&
	\includegraphics[width=4cm]{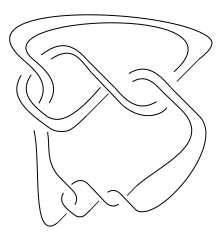}\\
	\mathrm{a.}\ \mathcal{S}_{Tor_{3}}^{Tw_{-2}}&\mathrm{b.}\ \mathcal{S}_{Tw_{-2}}^{Tw_{-2}}\\
	\end{array}
	$$
	\caption{Torus and twist satellites of the figure-eight knot.\label{fig:sat}}
\end{figure}

\paragraph{Two-strand pattern in satellites.}
A more complex case is that of two-strand (torus and twist) satellites (Fig.\ref{fig:sat}). In \cite{ALMSat}, we examined these satellites of the two-strand torus and twist knots, for prime knots up to 7 crossings, and for separate knots with 8--10 crossings, and for the torus knots $T[3,5]$ and $T[3,7]$. Now the $\Theta$-factor appears as a coefficient of one summand in the answer,
\be
(q^3t)^{-n}Kh^{S_{Tor_{2,n}}^{\mathcal K}}=Kh^{Tor_{2,n+s}}+(q^t)^{-n}\mu\mathfrak{K}^{\mathcal K},\qquad
Kh^{S_{Tw_{n}}^{\mathcal K}}=Kh^{Tw_{n+s}}+(q^t)^{-n}\tau\mathfrak{K}^{\mathcal K},
\label{KhSat}\ee
where $\mu=-t^{-1}\frac{1-q^6t^3}{1-q^2t}$, $\tau=q(1+q^2t)\mu$ are knot-independent, 
and $s$ is an integer invariant of $\mathcal{K}$.\footnote{We were not able to identify $s$ with any already known knot invariant.}

The $\Theta$-jumps in (\ref{KhSat}) are contained in $Kh^{Tor_{2,n+s}}$ (\ref{KhTor}), $Kh^{Tw_{n+s}}$ (\ref{KhTwist}), while the function $\mathfrak{K}$ is free of jumps in $n$. The boundary condition at $t=-1$ for (\ref{KhSat}) imply that $\mathfrak{K}$ only slightly differs from the coloured HOMFLY polynomial. Hence $\mathfrak{K}$ can be used as a jump-free substitute of the coloured $Kh$ polynomial \cite{ALMSat}.

\section{Weird catastrophes}

Below we give some examples of catastrophes that we call \textit{weird}, where jumps are not of type (\ref{KhTor}), and where the $KhR_N^{\prime\prime}$ in (\ref{gencat}) is not obviously present in the $KhR_N$.

\paragraph{``Thick'' pretzel knots}
Apart from the described ``regular'' regions\footnote{and the degenerated cases with $\{n_0,n_1,n_2\}\in\{0,\pm 1\}$ that we do not consider here}, there is the exceptional regions $-n_1<n_2<0<n_1\le n_0$ and $n_2<n_1<0<n_0<-n_1$.
The jump of the $KhR$ for the genus two pretzel knots near the boundary, e.g., of the first exceptional region looks like
\be
Kh^{P_{n_0,n_1,n_2}}=q^3\Phi^{\mathrm{exc}}_{n_0,n_1,n_2}(q^2t),\qquad
\Phi^{\mathrm{exc}}_{n_0,n_1,n_2}(\lambda)=\lambda^{n_0+n_1+n_2}
\left\{\lambda^{n_2}\Phi^{\mathrm{exc}(1)}(\lambda)+(-t)^{\Theta_{\mathrm{exc}}}\Phi^{\mathrm{exc}(2)}(\lambda)\right\}
\nonumber\\
\Phi^{\mathrm{exc}(1)}=
F_{n_0+n_2}\big(\textstyle{\frac{1}{\lambda}}\big)\Big\{F_{n_1+n_2-1}\big(\textstyle{\frac{1}{\lambda}}\big)+
\textstyle{\frac{1}{\lambda}}\Big\}
,\qquad 
\Phi^{\mathrm{exc}(2)}=f_{n_2-1}\big(\textstyle{\frac{1}{\lambda}}\big)F'_{n_2+1}\big(\lambda\big),
\label{KhPretzExc}\ee
where
$
f_n(\lambda)=\frac{1-\lambda^{n-1}}{1-\lambda},
$
and
$
\Theta_{\mathrm{exc}}=\left\{\begin{array}{l}1,\ -\min(n_1,n_2)<n_2<0<n_1,n_2\\
0,\ \text{otherwise}\end{array}\right..
$

Here we should point out two special properties of the pretzel knots in the exceptional regions.
First, the ``correction'' $\Phi^{\mathrm{exc}(2)}$ is the only term in $Kh^{P_{n_0,n_1,n_2}}$ that includes $n_2$ in two multiplies at the same time. In this sense\footnote{We called this phenomenon a nimble evolution in \cite{APev}.},  $\Phi^{\mathrm{exc}(2)}$ and $Kh^{P_{n_0,n_1,n_2}}$ in the exceptional regions depend on $(q^2t)^{2n_2}$ instead of $(q^2t)^{n_2}$. 

Second, all pretzel knots at the exceptional regions are ``homologically thick'', i.e., their $KhR_N$ are \textit{not} obtained from the corresponding HOMFLY just by substitution that preserves an analytic dependence on $N$ \cite{APev}. 
Accordingly, computer simulations with $KhoCa$ \cite{Lewark} (applicable when one $n_i$ is even) demonstrate the jump of $KhR_N$ as a function of $N$ between $N=2$ and $N=3$ 
(with analytic dependence for $N\ge 3$), both for pretzel knots of genus two and of higher genera.

\paragraph{Mirror symmetry for multi-strand and cabled braids}

In \cite{AMev}, the Khovanov polynomial for a positive three-strand torus knot was expressed as a function of the crossing number with the two branches,
\be
Kh^{Tor_{3,3n+p}}(q,t)=\mathcal{K}_{3,n}^{p}(q,t),\qquad p=1,2,\ n>0.
\ee
Although the Khovanov polynomial possesses mirror symmetry by construction,
\be
Kh^{Tor_{3,3n+1}}(q,t)=Kh^{Tor_{3,-3n-1}}\big(\textstyle{\frac{1}{q}},\textstyle{\frac{1}{t}}\big),
\ee
the function $\mathcal{K}$ does not. Instead
$
\mathcal{K}_{3,n}^{1}(q,t)=
-\frac{1}{t}\mathcal{K}_{3,n}^{2}\big(\textstyle{\frac{1}{q}},\textstyle{\frac{1}{t}}\big).
$
Starting from the four strands, similar formulas for positive torus knots $T[m,nm+p]$ ($1\le p \le m-1$ is relatively prime with $m$) contain a function $\mathcal{K}_{m,n}^{p}(q,t)$, which is sign indefinite for negative $n$. The $Kh$ of these knots are given by another analytic function $\widetilde{\mathcal{K}}_{m,n}^{p}(q,t)$, where, e.g., 
$
\widetilde{\mathcal{K}}_{4,n}^{1}(q,t)\not{\!\!\sim\ }\mathcal{K}_{4,n}^{-1}(\textstyle{\frac{1}{q}},\textstyle{\frac{1}{t}}),
$
and similarly for higher number of strands.
In addition,  the knots $Tor_{m,\mp1}$ with $n=0$ are the unknots with $Kh=1$, which is not the value of the corresponding analytic functions, nor for positive, nor for negative $n$.

We observed similar problem for those four-strand non-torus knots, which are two-strand satellites of the two-strand knots, as well as for two-strand satellites of twist knots and for twist satellites of both types of knots.
The explicit formulas for the positive and negative knots are given in \cite{AMPSat}, and they are indeed related non-trivially.
\paragraph{Eigenvalues for torus and cabled braids}
The Khovanov and even Khovanov-Rozansky polynomials of all torus knots include the dependence on the crossing number via the two-strand factor $f_n(\lambda)=\frac{1-\lambda^n}{1-\lambda}$. 
But the larger the number of strands becomes, the more there are different eigenvalues $\lambda$ \cite{AMev}. 
Namely, for some positive polynomials $P_I^{\pm}(q,t)$ (below $I\in\{\emptyset,1,2,12,13,23,24\}$),
\be
&\boxed{N=2}&\Lambda=-q^2t\\
&\lambda_1=q^4t^2&\mathcal{K}_{2,2n+1}=\Lambda^n\big\{1+P^{\pm}(q,t)f_n(\lambda_1)\big\}\nn\\
&\lambda_2=q^6t^4&\mathcal{K}_{3,3n\pm1}=\Lambda^n\big\{1+P^{\pm}_1(q,t)f_n(\lambda_2)\big\}\nn\\
&\lambda_3=q^8t^6&\mathcal{K}_{4,4n\pm1}=\Lambda^n\big\{1+P^{\pm}_1(q,t)\textstyle{\frac{1}{1-\lambda_1}}
\big(P^{\pm}_{12}(q,t)f_{2n}(\lambda_2)+P^{\pm}_{23}(q,t)f_n(\lambda_3)\big)\big\}\nn\\
&&n>0\nn\ee
But the number of eigenvalues also grows with the gauge group rank $N$ if $n>2$. E.g.,
\be
&&\boxed{N=3}\qquad 
\lambda_4=q^8t^5,\qquad _3\Lambda=-q^4t,
\\&&\!\!_3\mathcal{K}_{4,4n\pm1}\!=\!{_3\Lambda}^n\big\{1+P^{\pm}_2\textstyle{\frac{1}{(1-\lambda_1)^2}}f_{2n}(\lambda_2)
\!+\!P^{\pm}_1(q,t)\textstyle{\frac{1}{(1-\lambda_1)(1-\lambda_1^2)}}
\big(P^{\pm}_{13}(q,t)f_n(\lambda_3)\!+\!P^{\pm}_{24}(q,t)f_n(\lambda_4)\big)\big\}.
\nonumber\ee
The four-strand eigenvalues are also present in the $Kh$ polynomials for the two-strand torus and twist satellites of torus and twist knots \cite{AMPSat}.

Hence, it seems that the presence of exponentials other than $\lambda=q^2t$ (catastrophe in $\lambda$), and a non-trivial relation of the $Kh=KhR_2$ and $KhR_3$ (catastrophe in $N$), and a weird catastrophe (in the winding number $n$) related to the non-trivial mirror symmetry of the $KhR_N$ polynomial of the same knot have something to do with each other.

\section{Conclusion}
Based on the above examples, we distinguish two  kinds of the $KhR_N$ catastrophes w.r.t form of the jump in the analytic expression for $KhR_N$. 
The catastrophes we call regular cause the jump (\ref{gencat}) with the positive polynomials $KhR_N^{\prime}$ and $KhR_N^{\prime\prime}$ (which in this sense are present in the $KhR_N$ in the entire parameter space). Moreover, $KhR_N^{\prime\prime}$ looks like or is composed of $KhR_N$ polynomials for the two-strand braid, whose jump is dictated by positivity of the $KhR_N$ polynomial \cite{AMev,APev,ALMSat}. We suppose that a regular catastrophe happens when the $KhR_N$ complex contains a subcomplex for a positive two-strand braid, and it flips to a subcomplex for a negative two-strand braid \cite{DM3,AMKhov}. 

Other catastrophes that we call weird cause the jumps, where the $KhR^{\prime}_N$ and $KhR_N^{\prime\prime}$ in (\ref{gencat}) are not necessarily positive \cite{AMev,APev,ALMSat} (unlike the $KhR_N$). 
Moreover, the analytic dependence of $KhR_N$ on the parameters of the knot family can severely differ on both sides of the jump. Such jumps happen when analytic dependence contains exponentials ${\lambda_k}^{n_i}$ with $\lambda_k$ other than two-stand value $\lambda=q^2t$. E.g., new $\lambda_k$ come from the new two-stand factors $f_n(\lambda_k)$ in our higher-strand torus formulas and from the term $F_{n_2}(q^2t)F'_{n_2}(q^2t)\sim \left(q^4t^2\right)^{n_2}+\ldots$ in our formulas for the ``thick'' pretzel knots. The most surprising is that the Khovanov polynomials near a weird catastrophe have jumps in analytic dependence on $N$ for some $N=N_0$. Hence, we conjecture that there is a subcomplex (of the $KhR_N$ complex) that is not two-strand periodic and contains morphisms degenerating for some values of $N$. Shrinking of such subcomplex causes a weird catastrophe for the $KhR_N$.

Our next goal is then to study the $KhR_N$ complexes for the knot families  discussed here, especially when a catastrophe happens. We expect this to help us both in understanding the already observed phenomena and in making new predictions for more general knot families.

\section*{Ackhowledgements}
This work was supported by Basis foundation PostDoc-22-1-3-34-1.

\printbibliography

\end{document}